\documentclass[twocolumn,prl,showpacs]{revtex4}
\usepackage{graphicx}
\usepackage{dcolumn}
\usepackage{bm}
\usepackage{amsmath}

\begin{document}

\preprint{Helical edge}
\title{Topological spin-current in non-centrosymmetric
superconductors }
\author{ Yukio Tanaka$^{1}$, Takehito Yokoyama$^{1}$,
Alexander V. Balatsky$^{2}$ and Naoto Nagaosa$^{3,4}$}
\affiliation{$^1$Department of Applied Physics,
Nagoya University, Nagoya, 464-8603,
Japan \\
$^2$ Theoretical Division,
Los Alamos National Laboratory, Los Alamos, New Mexico 87545, USA \\
$^3$ Department of Applied Physics, University of Tokyo, Tokyo 113-8656, 
Japan \\
$^4$ Cross Correlated Materials Research Group (CMRG), ASI, RIKEN, WAKO 
351-0198, Japan
}
\date{\today}

\begin{abstract}
We study the spin transport properties of the non-centrosymmetric
superconductor with time-reversal-symmetry
where spin-triplet $(p_{x} \pm i p_{y})$-wave and spin-singlet $s$-wave 
pair potential can mix each other. 
We show that when the amplitude of 
$(p_{x} \pm ip_{y})$-wave pair potential is larger than 
that of $s$-wave one, the superconducting 
state belongs to the topologically nontrivial class analogous
to the quantum spin Hall system, and the resulting
helical edge modes as Andreev bound states 
are topologically protected. We find 
that the incident angle dependent spin polarized current flows
through the interface due to the presence of the helical edge modes.
With a weak magnetic field, also the angle-integrated
current is strongly spin polarized. 
\end{abstract}

\maketitle

%--- title ---

%--- author ---

%
%--- address ---

%
%--- date ---

% It is always \today, today,
% but any date may be explicitly specified
%-----------------------------------------------------------
% Abstract
%-----------------------------------------------------------

%-----------------------------------------------------------

% PACS, the Physics and Astronomy
% Classification Scheme.
%\keywords{Suggested keywords}%Use showkeys class option if keyword
%display desired
%\section{Introduction}
%-----------------------------------------------------------

The topological properties of the electronic states 
have been attracting intensive interests in condensed 
matter physics. Especially, it was highlighted by the 
discovery of the quantum Hall system (QHS) showing the 
accurate quantization of the Hall conductance $\sigma_H$ 
which is related to the topological integer \cite{Girvin,Thouless}. 

Recently, the concept of the QHS has been generalized to 
the time-reversal ($T$) symmetric system, 
i.e., the quantum spin Hall system 
(QSHS) \cite{Mele,Bernevig,Fu}. QSHS could be regarded as  
the two copies of QHS for up and down spins with 
the opposite chiralities. 
In generic case, however, the mixture of up and 
down spins occurs due to the spin-orbit interaction,
which necessitates the new topological number to characterize
QSHS \cite{Mele,Fu}. 
%and there is no quantization of 
%physical observables.
In QSHS, there exist the helical edge modes, i.e.,
the time-reversal pair of right- and left-going 
one-dimensional modes, which has been
experimentally demonstrated for the quantum well of HgTe system by
the measurement of the charge conductance \cite{Konig}. \par
%with the Kramers pair of Majorana edge modes.
%Also the Majorana fermion states 
%with the superconducting proximity effect at the surface of the 
%QSHS has been proposed \cite{FuKane}.%
%
In the field of superconductivity, 
the chiral $p$-wave superconductors such as Sr$_2$RuO$_4$ \cite{Maeno}
can be considered as an analogue of the 
QHS, and novel phenomena such as one-dimensional 
Majorana fermions (real fermions) modes at the edge 
\cite{MatsumotoSigrist,ReadGreen} 
and the non-Abelian statistics of the vortex 
\cite{ReadGreen,Ivanov} has been proposed there. 
Beside these issues, 
the non-centrosymmetric (NCS) superconductors such as CePt$_3$Si
are a central topic \cite{Bauer,Frigeri}. 
Also the two-dimensional NCS superconductors are 
expected at the interfaces and/or surfaces due to the 
strong potential gradient. An interesting example is
the superconductivity at LaAlO$_3$/SrTiO$_3$ interface \cite{Interface}. 
In NCS superconductors, the spin-orbit interaction comes into play. 
Especially, Frigeri et al. \cite{Frigeri} have shown that 
$(p_{x} \pm ip_{y})$-pairing 
state has the highest $T_c$ within the triplet-channel in CePt$_3$Si. 
However, the singlet ($s$-wave) and triplet ($p$-wave) 
pairings are mixed, and several novel associated properties 
such as the large upper critical field beyond the 
Pauli limit have been focused on \cite{Frigeri}. 
On the other hand, the pure $(p_{x} \pm ip_{y})$-pairing state 
has been studied from the viewpoint
of the superconducting analogue of QSHS \cite{Qi}.
Therefore, it is an important and urgent issue to study the 
spin transport properties of the NCS superconductors
from the topological viewpoint.

%and the $Z_2$ topological number is
%defined, which is related to the Kramers's doublet due to the
%$T$-symmetry . This $Z_2$ number is related to
%the spin transport associated with the adiabatic penetration of the
%$\pi$-magnetic flux, and also to the number of the helical mode
%pairs near the edge of the sample. When the number is odd, there
%always remain at least one pair, which is protected from the
%disorder scattering by the Kramers's theorem. This $Z_2$ number 
%is reduced to 
%the half of the spin Chern number (the difference between the spin
%up and spin down Chern numbers) mod 4 when these bands are decoupled
%\cite{Fu}. 

In this Letter, we study the spin transport properties of the
non-centrosymmetric (NCS) superconductor \cite{Bauer} with
$T$-symmetry, where 
$(p_{x} \pm i p_{y})$-wave and spin-singlet $s$-wave 
pair potential can mix each other. 
We show that when the amplitude of 
$(p_{x} \pm ip_{y})$-wave pair potential is larger than 
that of $s$-wave one, the superconducting 
state belongs to the topologically nontrivial class analogous
to the quantum spin Hall system, and the resulting
helical edge modes as Andreev bound states(ABS) 
are topologically protected.
We study Andreev reflection \cite{Andreev} at low energy, which is
determined mostly by the helical edge modes, and find the
incident angle dependent spin polarized current flowing through the
interface. When the magnetic field is applied, even the
angle-integrated current is spin polarized.

We start with the Hamiltonian of NCS superconductor
\[
\check{H} = \left( {\begin{array}{*{20}c}
{\hat H\left( {\bf k} \right)} & {\hat \Delta \left( {\bf k} \right)} \\
{ - \hat \Delta ^ * \left( { - {\bf k}} \right)} & { - \hat H^ * \left( 
{ - {\bf k}} \right)} \\
\end{array}} \right)
\]
with
$\hat{H}({\bm k})
=\xi_{{\bm k}} + \bm{V}(\bm{k}) \cdot \hat{\bm{\sigma}}
$,
$\bm{V}({\bm k})=\lambda (\hat{\bm{x}}k_{y}-\hat{\bm{y}}k_{x})
$,
$\xi_{\bm{k}}=\hbar^2 {\bm k}^{2}/(2m) - \mu$.
Here, $\mu$, $m$,
$\hat{\bm{\sigma}}$ and $\lambda$ denote
chemical potential, effective mass,
Pauli matrices and coupling constant of Rashba spin-orbit
interaction, respectively \cite{Frigeri}.
The pair potential $\hat{\Delta}(\bm{k})$
is given by 
\begin{equation}
\hat{\Delta}(\bm{k}) = 
[\bm{d}(\bm{k})\cdot \hat{{\bm \sigma}}]i\hat{\sigma}_{y}
+i \psi(\bm{k})\hat{\sigma}_{y}. 
\end{equation}
We choose $(p_{x} \pm ip_{y})$-wave pair for spin-triplet component 
with 
$\bm{d}(\bm{k})=\Delta_{p}(\hat{\bm{x}}k_{y}-\hat{\bm{y}}k_{x})/
\mid {\bm k}\mid$ 
\cite{Frigeri} 
and 
$s$-wave one with
$\psi(\bm{k})=\Delta_{s}$ with $\Delta_{p} \geq 0$ and 
$\Delta_{s} \geq 0$. 
The superconducting gaps 
$\Delta_{1}=\Delta_{p}+\Delta_{s}$ and 
$\Delta_{2}=\mid \Delta_p-\Delta_s \mid$ open for the 
two spin-splitted band, respectively, in the homogeneous 
state \cite{Iniotakis}. \par
However, as seen below, surface states 
are crucially influenced by the relative magnitude 
between $\Delta_{p}$ and $\Delta_{s}$.
% which depends on the 
%form of the inter-electron potential $V_{k, k'}$.  
%%%%%%%%%%%%%%%%%%%
%We need this or something like this as you make connection
%to adiabatic turn on of lambda in discussion on Chern index*** 
%%%%%%%%%%%%%%%
Let us  consider wave function including ABS localized at the
surface. 
Consider a two-dimensional semi-infinte superconductor
on $x>0$ where the surface is located at $x=0$. 
The corresponding wave function 
%
%surface localized edge state
is given by \cite{Yokoyama}
\begin{eqnarray}
\Psi_{S}(x)=\exp(ik_{y}y)
[c_{1}\psi_{1}\exp(iq^{+}_{1x}x) + c_{2}\psi_{2}\exp(-iq^{-}_{1x}x)
\nonumber\\
+ c_{3}\psi_{3}\exp(iq^{+}_{2x}x) + c_{4}\psi_{4}\exp(-iq^{-}_{2x}x) ], 
\\
q^{\pm}_{1(2)x}
=k^{\pm}_{1(2)x} \pm \frac{ k_{1(2)}}{k^{\pm}_{1(2)x}}
\sqrt{\frac{E^{2}-\Delta_{1(2)}^{2}}
{\lambda^{2} + 2\hbar^{2}\mu /m }}, 
\nonumber
\end{eqnarray}
with 
$k^{+}_{1(2)x}=k^{-}_{1(2)x}=k_{1(2)x}$ for $\mid k_{y} \mid \leq k_{1(2)}$ 
and $k^{+}_{1(2)x}=-k^{-}_{1(2)x}=k_{1(2)x}$ for $\mid k_{y} \mid >k_{1(2)}$.  
Here, $k_{1}$ and $k_{2}$ are Fermi momentum of the small and large
magnitude of Fermi surface given by $-m\lambda/\hbar^{2} + \sqrt{
(m\lambda/\hbar^{2})^{2} + 2m\mu/\hbar^{2} }$ and $m
\lambda/\hbar^{2} + \sqrt{ (m\lambda /\hbar^{2})^{2} +
2m\mu/\hbar^{2} }$, respectively. $k_{1(2)x}$ denotes the $x$
component of the Fermi momentum $k_{1(2)}$, with
$k_{1(2)x}=\sqrt{k_{1(2)}^{2} -k_{y}^{2}}$. The wave functions are
given by $^T\psi_{1} =\left(
u_{1},-i\alpha_{1}^{-1}u_{1},i\alpha_{1}^{-1}v_{1},v_{1} \right)$,
$^T\psi_{2} =\left( v_{1},-i\tilde{\alpha}_{1}^{-1}v_{1},
i\tilde{\alpha}_{1}^{-1}u_{1},u_{1} \right)$, 
$^T\psi_{3} =\left(u_{2},i\alpha_{2}^{-1}u_{2},
i\gamma \alpha_{2}^{-1}v_{2},-\gamma v_{2} \right)$, 
and $^T\psi_{4} =\left( v_{2},i\tilde{\alpha}_{2}^{-1}v_{2}, 
i\gamma \tilde{\alpha}_{2}^{-1}u_{2},-\gamma u_{2} \right)$, 
with $\gamma = \rm{sgn}(\Delta_{p}-\Delta_{s})$. 
%%For the $s$-wave superconductor, on the
%other hand, $^T\psi_{1} =\left(
%u_{0},-i\alpha_{1}^{-1}u_{0},i\alpha_{1}^{-1}v_{0},v_{0} \right)$,
%$^T\psi_{2} =\left( v_{0},-i\tilde{\alpha}_{1}^{-1}v_{0},
%i\tilde{\alpha}_{1}^{-1}u_{0},u_{0}\right)$, $^T\psi_{3} =\left(
%u_{0},i\alpha_{2}^{-1}u_{0},-i\alpha_{2}^{-1}v_{0},v_{0}\right)$,
%and $^T\psi_{4} =\left( v_{0},i\tilde{\alpha}_{2}^{-1}v_{0},
%%-i\tilde{\alpha}_{2}^{-1}u_{0},u_{0}\right)$. 
In the above, $u_{1(2)}$ and  $v_{1(2)}$  
are given as $\sqrt {\frac{1}{2}\left( {1 + \frac{{\sqrt
{E^2 - \Delta_{1(2)} ^2 } }}{E}} \right)}$, and $\sqrt
{\frac{1}{2}\left( {1 - \frac{{\sqrt {E^2 - \Delta_{1(2)} ^2 } }}{E}}
\right)}$. 
%, respectively, 
%with $\Delta_{1}=\Delta_{p} + \Delta_{s}$ and 
%$\Delta_{2}= \mid \Delta_{p} -\Delta_{s} \mid$. 
Here we have introduced
$\alpha_{1}=(k^{+}_{1x}-ik_{y})/k_{1}$,
$\alpha_{2}=(k^{+}_{2x}-ik_{y})/k_{2}$,
$\tilde{\alpha}_{1}=(-k^{-}_{1x}-ik_{y})/k_{1}$, and
$\tilde{\alpha}_{2}=(-k^{-}_{2x}-ik_{y})/k_{2}$.  
%with $\tilde{k}_{1x}=k_{1x}$ for $k_{1} \geq \mid k_{y} \mid$ 
%and $\tilde{k}_{1x}=-k_{1x}$ for $k_{1} < \mid k_{y} \mid$. 
$E$ is the
quasiparticle energy measured from the Fermi energy.
\par
%%The wave functions for $p_{x} \pm ip_{y}$-wave superconductor
%and $s$-wave one are almost the same \cite{Yokoyama}.
%The only difference is the
%sign of third and fourth column of $\psi_{3}$ and $\psi_{4}$.
%%However, the resulting bound state solution is very different.
By postulating $\Psi_{S}(x)=0$ at $x=0$, we can determine the
ABS. 
The bound state condition can be expressed by 
\begin{eqnarray}
\sqrt{(\Delta_{1}^{2}-E^{2})(\Delta_{2}^{2}-E^{2})}
=\frac{1 -\zeta}{1 + \zeta}(E^{2} + \gamma \Delta_{1}\Delta_{2}), 
\label{bound}
\\
%\begin{equation}
\displaystyle
\zeta =
\left\{
\begin{array}{ll}
\frac{\sin^{2}[\frac{1}{2}(\phi_{1} + \phi_{2})]}
{\cos^{2}[\frac{1}{2}(\phi_{1}-\phi_{2})]}
& \mid \phi_{2} \mid \leq \phi_{C} \\
1 & \phi_{C} < \mid \phi_{2} \mid \leq \pi/2, 
\end{array}
\right.
\end{eqnarray}
%\end{equation}
with $\zeta \leq 1$, 
$\cos\phi_{1}=k_{1x}/k_{1}$ and $\cos\phi_{2}=k_{2x}/k_{2}$.
The critical angle $\phi_{C}$ is defined as
$\sin^{-1}(k_{1}/k_{2})$. 
For $\lambda=0$, eq. (\ref{bound}) reproduces the 
previous result \cite{Iniotakis}. 
As seen from eq. (\ref{bound}), the ABS including 
zero energy state is only possible 
for $\mid \phi_{2} \mid \leq \phi_{C}$ and 
$\gamma=1$, $i.e.$, $\Delta_{p} > \Delta_{s}$.  
The present ABS is just the edge state, where the
localized quasiparticle can move along the edge. 
The energy level of the edge state depends crucially
on the direction of the motion of the quasiparticle.
%
%{\bf{ Tanaka-sensei: This sentence is not so clear to me. The position means th%e energy or real space position ?}} 
The inner gap edge modes are absent for large magnitude of $k_{y}$, $i.e.$
$\phi_{2}$. 
%In this case, $k_{1x}$ becomes pure imaginary number due
%to the conservation of the Fermi momentum parallel to the surface.
The parameter regime where the edge modes survive is reduced with
the increase of the magnitude of $\lambda$. 
However, as far as we concentrate on the perpendicular injection,
the edge modes survive as 
the mid gap ABS \cite{ABS,TK95}  irrespective of the
strength of $\lambda$. If we focus on the low energy limit, 
ABS can be written as
\begin{eqnarray}
E=\pm \Delta_{p}(1 - \frac{\Delta^{2}_{s}}{\Delta^{2}_{p}})
\frac{k_{1} + k_{2}}{2k_{1}k_{2}}k_{y}, 
\end{eqnarray}
with $\Delta_{s}<\Delta_{p}$
for any $\lambda$ with small magnitude of $k_{y}$. 
%%%%%%%%%%%%%%%%%%%%%%%%%%%%%%%%%%%%#
%ここで（３）式との関係をレフェリーAが要求しているようにもう少し具体的に議論した%方が良いと思います。\Delta_2, \gammaの符号が両方変るので、（３）式に符号変化が%%現れないように思ってしまうのですが、、、。
%%%%%%%%%%%%%%%%%%%%%%%%%%%%%%%%%%%%%%%%%%%%%%
For $\Delta_{s} \geq \Delta_{p}$, the presnet ABS 
vanishes since the value of right side of eq. (3) becomes negative 
due to the negative sign of $\gamma$ 
for $\mid E\mid<\Delta_{1}$ and 
$\mid E \mid<\Delta_{2}$. 
It should be remarked that the present ABS do not break the 
time reversal symmetry, since the edge current carried by each
Kramers doublet flows in the opposite direction.
Thus they can be regarded as helical edge modes, where 
two modes are connected to each other by time reversal operation. 

Now we give an argument why the superconducting state 
with $\Delta_{p}>\Delta_{s}$ has the ABS 
from the viewpoint of $Z_2$ (topological) class \cite{Mele}.
We commence with the pure $(p_{x} \pm i p_{y})$-wave state 
without the spin-orbit interaction $\lambda$. 
Spin Chern number \cite{Fu} for the Bogoliubov-de Gennes (BdG) Hamiltonian 
is 2. 
Turning on $\lambda$ adiabatically, which leaves the 
$T$-symmetry intact and keeps the gap open, 
one can arrive at the BdG Hamiltonian of interest. 
Upon this adiabatic change of $\lambda$, the number of 
the helical edge mode pairs
% i.e., the $Z_2$-number, 
does not change. 
Then we increase the magnitude of $\Delta_{s}$ from zero. 
As far as $\Delta_{p} > \Delta_{s}$ is satisfied, 
the number of helical edge modes 
%the 
%$Z_{2}$ number 
does not change, since it is a topological number. 
However, if $\Delta_{s}$ exceeds $\Delta_{p}$, the helical mode 
disappears. In this regime, the topological nature of 
superconducting state belongs 
to pure $s$-wave state without $\lambda$.  
It is remarkable, just at $\Delta_{s}=\Delta_{p}$, one of the 
energy gap of the quasiparticle in the bulk closes,
where a quantum phase transition occurs. 
\par
%Chern number remains 1 and 0 for
%the $p_{x} \pm ip_{y}$ and $s$-pairing state, respectively. \par
%Note that one can continue from the $s$- to the $p_{x} \pm
%ip_{y}$-pairing state without closing the gap through the
%Hamiltonian by {\it{breaking}} the $T$-symmetry. 
%\par
%
%However, the
%$Z_2$-number can not be defined once $T$-symmetry is broken, and the
%adiabatic argument can not be used any more.( do we need this
%statement here. it seem tangential for the discussion).

Now we turn to the spin transport property governed by 
the ABS in the NCS superconductors \cite{Eschrig}.
First, we point out that the spin Hall effect, i.e., the 
appearance of the spin Hall voltage perpendicular to 
the superconducting current is suppressed by the 
compressive nature of the superconducting state 
by the factor of $( k_F \lambda_m)^{-2}$
($k_F$: Fermi momentum, $\lambda_m$: penetration depth) 
\cite{Furusaki}.  
Instead, we will show below that the 
spin transport through the junction between 
the ballistic normal metal at $x<0$ and NCS superconductor,
i.e., (N/SC) junction, can be enhanced by the Doppler
effect at the Andreev reflection.
We assume an insulating barrier at $x=0$
expressed by a delta-function potential $U \delta(x)$. 
%%%%%%%%%%%%%%%%%%%%%%%%%%%%%%%%%%%%%%%%%%%%%%%%%%%%%%%%%%%%%%%%%%%%%%
The wave function for spin $\sigma$ in
the normal metal $\Psi_{N}(x)$ 
%and in the NCS superconductor 
%$\Psi_{S}(x)$
is given by
\begin{eqnarray}
\Psi_{N}(x)\!\!&=&\!\!\exp(ik_{Fy}y)
[(\psi_{i\sigma}+\sum_{\rho=\uparrow,\downarrow}a_{\sigma,\rho}\psi_{a\rho})\exp(ik_{Fx}x)
\nonumber\\
&&+ \sum_{\rho=\uparrow,\downarrow}
b_{\sigma,\rho}\psi_{b\rho}\exp(-ik_{Fx}x)]
\end{eqnarray}
with
%$
%k_{Fx}^{\pm}=k_{Fx} \pm 
%\frac{\sqrt{E^{2}-\Delta_{p}^{2}}m}{\hbar^{2}k_{Fx}}$,
$^T\psi_{i\uparrow}$$=$
$^T\psi_{b\uparrow}$
$=$$\left(1,0,0,0 \right)$, 
$^T\psi_{i\downarrow}$=$^T\psi_{b\downarrow}$
$=$$\left(0,1,0,0 \right)$, 
$^T\psi_{a\uparrow}$
$=$$\left(0,0,1,0 \right)$, and 
$^T\psi_{a\downarrow}$
$=$$\left(0,0,0,1 \right)$. 
The corresponding $\Psi_{S}(x)$ is given by eq. (2). 
The coefficients $a_{\sigma,\rho}$ and $b_{\sigma,\rho}$ are 
determined by postulating the boundary condition between 
$\Psi_{N}(x)$ and $\Psi_{S}(x)$ given by 
$\Psi_{N}(0)=\Psi_{S}(0)$, and 
$\frac{d}{dx}\Psi_{S}(0)-\frac{d}{dx}\Psi_{N}(0)
=\frac{2mU}{\hbar^{2}}\Psi_{S}(0)$. 
We assume that the Fermi momentum $k_{F}$ of
normal metal and NCS superconductor before putting 
$\lambda$ are the same.
The quantities of interest are the angle resolved spin
conductance $f_{S}(\phi)$ and charge conductance $f_{C}(\phi)$
defined by \cite{Kashiwaya99}
\begin{eqnarray}
f_{S}(\phi)=[(\mid a_{\uparrow,\uparrow} \mid^{2}
- \mid a_{\uparrow,\downarrow} \mid^{2}
- \mid b_{\uparrow,\uparrow} \mid^{2}
+ \mid b_{\uparrow,\downarrow} \mid^{2})
\nonumber\\
-(\mid a_{\downarrow,\downarrow} \mid^{2}
- \mid a_{\downarrow,\uparrow} \mid^{2}
- \mid b_{\downarrow,\downarrow} \mid^{2}
+ \mid b_{\downarrow,\uparrow} \mid^{2})]\frac{\cos \phi}{2}, 
\nonumber\\
f_{C}(\phi)
=[2 + \sum_{\sigma,\rho} (\mid a_{\sigma,\rho} \mid^{2} -
\mid b_{\sigma,\rho} \mid^{2} ) ]
\frac{\cos \phi}{2}, 
\end{eqnarray}
where $\phi$ denotes the injection angle measured from the 
normal to the interface. 
First we consider pure $(p_{x} \pm i p_{y})$-wave state. 
In Fig. 1, the
angle resolved spin conductance is plotted as a function of
injection angle $\phi$ and bias voltage $V$ with $E=eV$.  
Note here that the $k_y$ is related to 
$\phi$ as 
$k_{y} = k_{F} \sin \phi$. 
It is remarkable that spin conductance has a non zero value although the
NCS superconductor does not break time reversal symmetry. 
$f_{S}(\phi)$ has a peak 
when the angle $\phi$ or $k_y$ 
gives the energy $E$ in the energy dispersion of ABS. 
With this condition, the spin-dependent Andreev reflection 
occurs to result in the spin current. 
%
%corresponding to the ABS as a helical edge mode. 
%which is given by $E=\Delta_{p}
%\sin[\frac{1}{2}(\phi_{1}+\phi_{2})]/
%\cos[\frac{1}{2}(\phi_{1}-\phi_{2})]$ obtained from eq. (3). 
Besides this property, we can show that
$f_{S}(\phi)=-f_{S}(-\phi)$ is satisfied. By changing the sign of
$eV$, $f_{S}(\phi)$ changes sign as seen in Fig. 1(a). 
Next, we look at the case where $s$-wave component coexists. 
We can calculate spin current similar to the pure 
$(p_{x} \pm i p_{y})$-wave case. 
For $\Delta_{s} < \Delta_{p}$, where helical edge modes exist, 
$f_{S}(\phi)$ has a sharp peak and 
$f_{S}(\phi)=-f_{S}(-\phi)$ is satisfied [see Fig. 1(b)]. 
These features are similar to those of pure
$(p_{x} \pm i p_{y})$-wave case. 
On the other hand, for  
$\Delta_{s} > \Delta_{p}$, where the helical edge modes are absent, 
sharp peaks of $f_{S}(\phi)$ as shown in Fig. 1 are absent. 
\par
%%%%%%%%%%%%%%%%%%%%%%%%%%%%%%%%%%%%%%%%%%%%%%%%%%%%%%%%%%%%%%%%%%%%%%
We have checked that there is negligible quantitative change by taking 
$\lambda=0$ limit compared to Fig. 1, $e.g.$, less 
than 0.5\% change of the peak height. 
%as long as the system belongs to the $Z_2$ nontrivial state. 
In this limit, for pure $(p_{x} \pm i p_{y})$-wave state,
$f_{S}(\phi)$ is given simply as follows 
\[
\frac{-8\sigma_{N}^{2}(1-\sigma_{N})\sin2\phi \sin2\varphi \cos\phi}
{\mid 4(\sin^{2}\phi - \sin^{2}\varphi) 
+ \sigma_{N}[\exp(-2i\varphi)(\sigma_{N}-2) 
+ 2 \cos2\phi] \mid^{2}},
\label{spincurrent}
\]
for $ \mid E \mid< \Delta_{p}$ and $f_{S}(\phi)=0$ for 
$ \mid E \mid > \Delta_{p}$ with 
$\sin \varphi = E/\Delta_{p}$
Transparency of the interface 
$\sigma_{N}$ is given by $4\cos^{2} \phi /(4\cos^{2} \phi + Z^{2})$
with a dimensionless constant $Z=2mU/\hbar^{2}k_F$.
The magnitude of $f_{S}(\phi)$ is largely enhanced at $E= \pm \Delta_{p}\sin\phi$ corresponding to the energy dispersion of ABS.
The origin of nonzero $f_{S}(\phi)$ even without $\lambda$ is due to the 
spin-dependent ABS. 
%As seen from eqs. (\ref{spincurrent})
%$f_{S}(\phi)$ becomes 
%zero for $\mid E \mid >\Delta_{p}$, since $\Gamma$ becomes a real number. 
%$f_{S}(\phi)$ becomes nonzero only for inner gap energy 
%$\mid E \mid < \Delta_{p}$. 
We have checked that even if  we take into account the 
spatial dependence of the $(p_{x} \pm i p_{y})$-wave 
pair potential explicitly, 
the resulting $f_{S}(\phi)$ does not qualitatively change \cite{Eschrig}. \par
Summarizing these features, we can conclude that the presence of the helical
edge modes in NCS superconductor is the
origin of the large angle resolved spin current through
N/NCS superconductor junctions. However, the magnitude of the angle
averaged normalized spin conductance becomes zero since
$f_S(\phi)=-f_S(-\phi)$ is satisfied. 
\par
%%%%%%%%%%%%%%%%%%%%%%%%%%%%%%%%%%%%%%%%%%%%%%%%%%%%%%%%%%
%CHANGE
%%%%%%%%%%%%%%%%%%%%%%%%%%%%%%%%%%%%%%%%%%%%%%%%%%%%%%%%
\begin{figure}[htb]
\begin{center}
\scalebox{0.8}{
\includegraphics[width=10cm,clip]{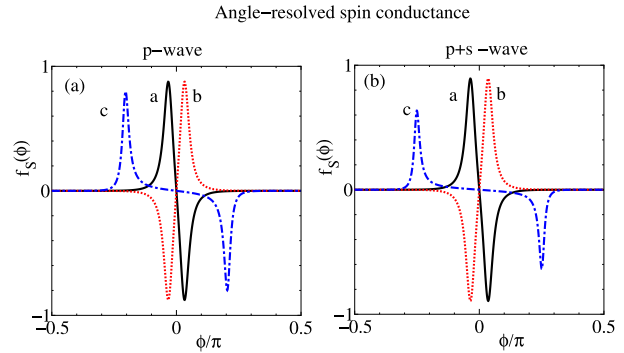}
}
\end{center}
\caption{(Color online) Angle resolved spin conductance for $Z=5$.
a: $eV=0.1\Delta_{p}$, b: $eV=-0.1\Delta_{p}$
and
c: $eV=0.6\Delta_{p}$ with $\lambda k_{F}=0.1\mu$. 
(a)pure $(p_{x}\pm ip_{y})$-wave case with $\Delta_{s}=0$.
(b)$\Delta_{s}=0.3\Delta_{p}$. }
\label{fig:1}
\end{figure}
Magnetic field offers an opportunity to observe the spin current in a 
much more accessible way, where $T$-symmetry is broken by the 
shielding current at the interface. Here
we consider the angle averaged normalized spin conductance
$\sigma_{S}$ and charge conductance $\sigma_{C}$ 
as a function of 
magnetic field which are given by \cite{TK95,Kashiwaya99}
\begin{equation}
\sigma_{S}=
\frac{\int^{\pi/2}_{-\pi/2} f_{S}(\phi) d\phi }
{\int^{\pi/2}_{-\pi/2} f_{NC}(\phi) d\phi}, \ \
\sigma_{C}=
\frac{\int^{\pi/2}_{-\pi/2} f_{C}(\phi) d\phi }
{\int^{\pi/2}_{-\pi/2} f_{NC}(\phi) d\phi}, 
\end{equation}
where $f_{NC}(\phi)$ denotes the angle resolved 
charge conductance in the normal state with 
$\Delta_{p}=\Delta_{s}=0$. 
Now we consider the magnetic field $H$ applied 
perpendicular to the two-dimensional plane, which induces
a shielding current along the N/NCS superconductor interface.
When the penetration depth of the NCS superconductor
is much longer than coherence length, the vector potential
can be approximated as
$\bm{A}(\bm{r})=(0,A_{y}(x),0)$ with
$A_{y}(x) = -\lambda_{m} H \exp(-x/\lambda_{m})$ 
with the penetration depth $\lambda_{m}$.
Here we consider the situation where 
the quantization of the Landau level can be neglected. 
Then quasiclassical approximation becomes available. 
The applied magnetic field shifts the quasiparticle energy 
$E$ in wave function of $\Psi_{S}(x)$ 
to $E-H\Delta_{p}\sin \phi/H_{0}$ with
$H_{0}=h/(2e\pi^{2}\xi \lambda_{m})$
and $\xi=\hbar^{2} k_{F}/(\pi m \Delta_{p})$ \cite{Doppler}. 
For typical values of $\xi \sim 10$nm, $\lambda_{m} \sim 100$nm,
the magnitude of $H_0$ is of the order of 0.2Tesla.
Here the order of the 
energy of Doppler shift is given by  $H \Delta_{p}/H_{0}$. 
Since the Zeeman energy is given by 
$\mu_{B}H$, the order of the 
energy of Doppler shift is 
$k_{F}\lambda_{m}$ times larger than that of Zeeman energy.  
Thus, we can neglect the Zeeman effect in the present analysis.
This is in sharp contrast to QSHS where the 
Zeeman energy is the main effect of $H$, which opens the gap in the 
helical edge modes and modulates the transport properties 
\cite{Konig}. The enhanced spin current due to the Doppler 
shift is specific to superconducting state not realized in QSHS. \par
\begin{figure}[htb]
\begin{center}
\scalebox{0.8}{
\includegraphics[width=10.0cm,clip]{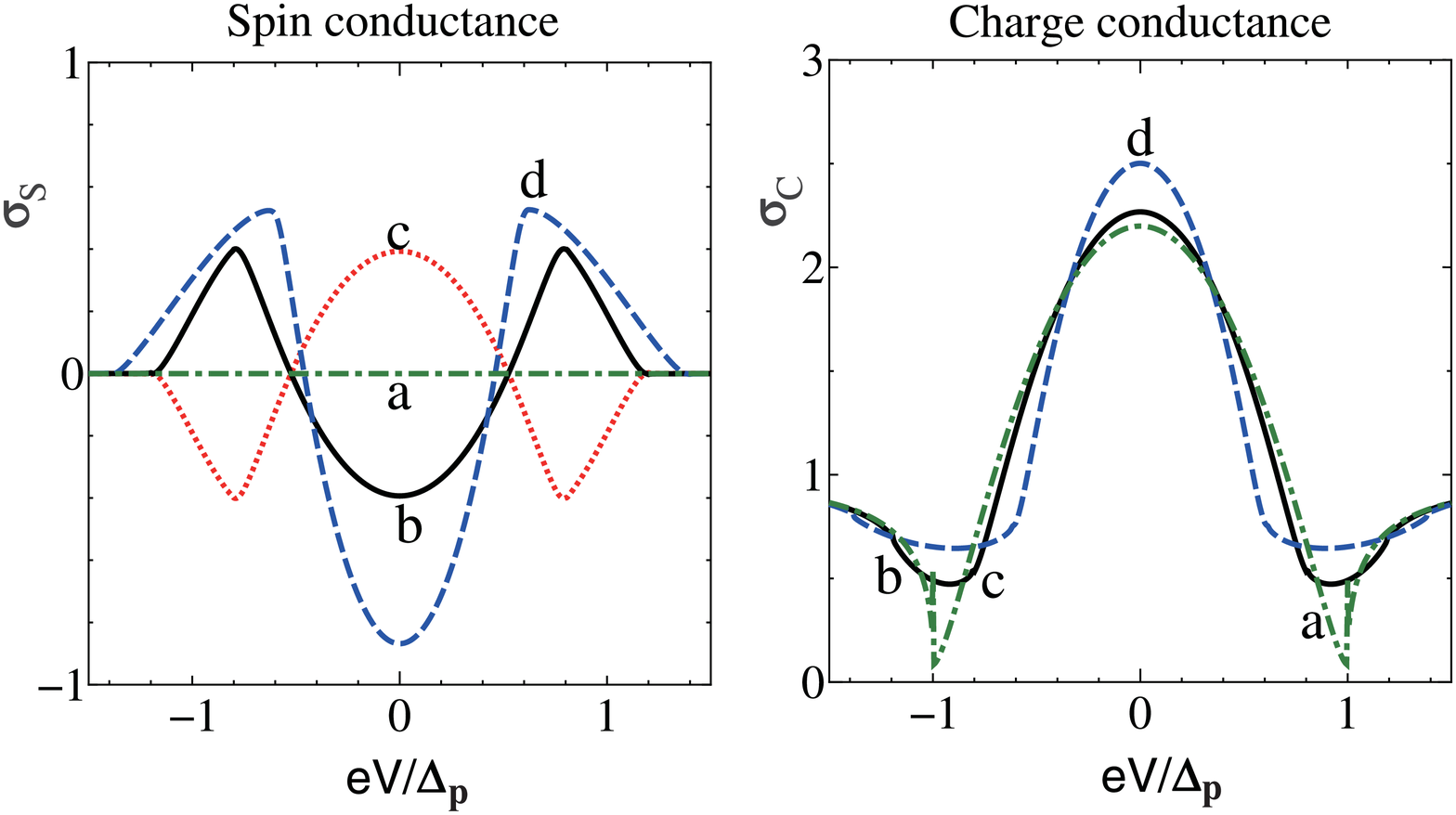}
}
\end{center}
\caption{(Color online) Angle averaged
spin conductance and charge conductance
as a function of $eV$ with bias voltage $V$ with 
$\lambda k_{F}=0.1\mu$.
a: $H=0$, b: $H=0.2H_{0}$, c:
$H=-0.2H_{0}$, and d: $H=0.4H_{0}$. Curves $b$ and $c$ of the right panel 
are identical.
}
\label{fig:2}
\end{figure}
As shown in Fig. 1, to discuss topological nature of the helical edge modes, 
it is sufficient to consider pure $(p_{x} \pm i p_{y})$-wave state. 
In the following, we choose $(p_{x} \pm i p_{y})$-wave case.
In Fig. 2, the spin conductance $\sigma_{S}$ and charge conductance
$\sigma_{C}$ normalized by the charge conductance in
the normal state are plotted. 
It should be noted that
$\sigma_{S}$ becomes nonzero in the presence of the magnetic field
$H$ (see curves $b$, $c$ and $d$), since  $f_{S}(\phi)$ 
is no more odd function of 
$\phi$ due to the imbalance of the helical edge modes. 
For $\lambda=0$ limit, 
the corresponding 
helical edge modes are given  by 
$E= \Delta_{p}(1 + H/H_{0})\sin\phi$ and $E= -\Delta_{p}(1 - H/H_{0})\sin\phi$. As seen from the curves $b$ and $c$, the
sign of $\sigma_{S}$ is reversed by changing the direction of the
applied magnetic field. 
%It should be remarked that the 
%emergence of the substantial magnitude of 
%spin conductance is due to the Doppler effect specific to 
%the superconductivity. 
On the other hand, the resulting charge
conductance has different features. For $H=0$, the resulting line
shape of $\sigma_{C}$ is the same as that of chiral $p$-wave
superconductor (see curve $a$ of right panel)
\cite{Yokoyama,Iniotakis,Linder}. 
%The wide zero bias conductance peak
%originates from the fact that the position of the ABS depends on the
%injection angle $\phi$ given by $\Delta_{p}\sin \phi$ \cite{Chiral}.
%The amplitude of $\sigma_{C}$ increases gradually with $H$. 
As seen
from curves $b$ and $c$ of right panel, $\sigma_{C}$ does not change
with the change of the direction of the magnetic field $H$. 
\begin{figure}[htb]
\begin{center}
\scalebox{0.8}{
\includegraphics[width=5cm,clip]{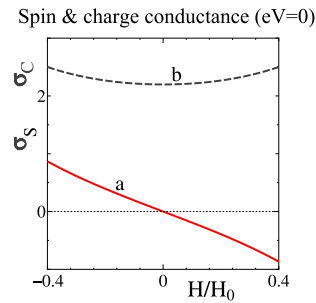}
}
\end{center}
\caption{(Color online) Angle averaged
spin and charge conductance
for $eV=0$ as a function of $H$ with $\lambda k_{F}=0.1\mu$. 
a: spin conductance and b: charge conductance.}
\label{fig:3}
\end{figure}
Finally, we show in Fig. 3 the zero-voltage $\sigma_{S}$ and $\sigma_{C}$.
$\sigma_{S}$ is nearly linearly proportional to $H$. 
Note that with a small magnetic field $H \cong 0.4H_0 \sim 100 {\rm Oe}$,
$\sigma_S$ is already of the order of 1.
Meanwhile, $\sigma_{C}$ is almost independent of $H$. 
%We hope
%the present dramatic behavior of $\sigma_{S}$ will be observed in
%actual experiments. 

In conclusion, we have studied the
spin transport property of non-centrosymmetric (NCS) 
superconductor from the viewpoint of topology and 
Andreev bound state (ABS). We have
found the incident angle dependent spin polarized current flowing
through the interface. When the weak magnetic field is applied, even the
angle-integrated current is largely spin polarized. 
As the analogy to quantum spin Hall system (QSHS), 
the ABS in NCS superconductor corresponds to the 
helical edge modes in QSHS. 
The Andreev reflection via helical edge modes 
produces the enhanced spin current specific to NCS superconductor. \par
%Y.T. was supported by Grant-in-Aid for Scientific Research (Grant No.
%20654030) %on Priority
%Area "Novel Quantum Phenomena Specific to Anisotropic Superconductivity"
%(Grant No. 17071007) and B (Grant No. 17340106)
%from the Ministry of Education, Culture, Sports, Science and Technology of
%Japan and NTT basic research laboratories.
%N.N. was partly supported by the
%Grant-in-Aids from under the Grant No.\ 15104006, No.\ 16076205, 
%No. 17105002, No.\ 19048015, and NAREGI Nanoscience Project from 
%the Ministry of Education, Culture, Sports, Science, and Technology, 
%Japan.
This work is partly supported by the
Grant-in-Aids from under the Grant No.\ 20654030, and
NAREGI Nanoscience Project from 
the Ministry of Education, Culture, Sports, Science, and Technology, 
Japan, NTT basic research laboratories, 
DOE BES and by LDRD.

%---------------------

\end{document}